# Resonant switching using spin valves.

K. Rivkin, J. B. Ketterson and W. Saslow

**Using micromagnetics we demonstrate that the r.f. field produced by a spin valve can be used to reverse the magnetization in a magnetic nanoparticle. The r.f. field is generated using a current that specifically excites a uniform spin wave in the spin valve. This current is swept such that the chirped-frequency generated by the valve matches the angular dependent resonant frequency of the anisotropy-dominated magnetic nanoparticle, as a result of which the magnetization reversal occurs. The switching is fast, requires currents similar to those used in recent experiments with spin valves, and is stable with respect to small perturbations. This phenomenon can potentially be employed in magnetic information storage devices or recently discussed magnetic computing schemes.**

**Micromagnetics, spintronics, ferromagnetic resonance**

Recently it has been shown that a logical element[1] can be constructed from a group of interacting magnetic nanoparticles, typically separated by no more than 50nm[2]. Behavior of the whole complex of nanoparticles is determined by the magnetization of a few "input" nanodots, which subsequently couple to the outside world. It was also proposed that spin valves can be used as such "input" elements, with the direction of their magnetization being determined by the applied current. The magnetization reversal of neighboring elements is achieved by producing strong magnetic fields that would exceed the effective anisotropy field in a neighboring magnetic dot and thereby force the realignment of the magnetization. Although novel, this method raises the following concern: one is forced to use materials with relatively small anisotropy, which means that the dots have to be relatively large to be thermally stable; the switching typically involves a non-uniform and possibly even chaotic precession of the magnetization, as a result of which the switching speeds can be relatively slow. We have earlier shown[3] that one can reverse the magnetization in a magnetic nanodot by using r.f. fields with a time-dependent frequency, so called "chirped" pulses, in this case one can use intrinsic resonant properties of the sample in order to facilitate a fast magnetization reversal with relatively small fields. The analytical and numerical analysis was carried out in the limit of a macrospin model when the whole system can be represented by just a single dipole in the presence of anisotropy and external r.f. field. In the present work we examine whether such switching can be accomplished by using a spin valve to generate the required r.f. field. Since in this case the distance between the spin valve and the magnetic nanodot is comparable to their dimensions, a micromagnetic analysis will be used. It is possible that such a mechanism of magnetization reversal can be used in magnetic logical elements and, possibly, lead to a new information storage technology.

We imagine a standard spin valve configuration similar to the one used by Kiselev et al.[4], consisting of two magnetic layers, separated by a nonmagnetic layer. One of the layers is assumed to have a constant magnetization, and the other, the so-called free layer, has a magnetic moment that can change with time. Initially both layers are magnetized along the same direction; the spin polarized current entering the free layer is also polarized along the same direction. Here we will use dimensions similar to those reported by Kiselev et al[5] in which the first (constant magnetization) layer is 20nm thick and the free layer is 2nm thick. The layers are formed from cobalt and approximate an ellipse with principal diameters 130 by 70nm. We assume an exchange stiffness $A = 1.3 \cdot 10^{-6}$ erg/cm and for a saturation magnetization $M_s = 795$ emu/cm$^3$ characteristic of permalloy; it is further assumed that there is no exchange interaction between the layers. In order to describe the precession of the magnetization in the free layer one can use the Landau-Lifshitz[6] equation in the presence of a dissipative Landau-Lifshitz term and a spin transfer torque[7]-[11]:

$$\frac{d\mathbf{m}}{dt} = -\gamma \mathbf{m} \times \mathbf{h} ; \qquad (1)$$

where

$$\mathbf{h} = \mathbf{h}^{true} + \frac{\mathbf{m}}{M_s} \times \left(\beta \mathbf{h}^{true} - I\mathbf{h}^J\right); \qquad (2)$$

here $\mathbf{m}$ is the magnetic moment and $\gamma$ is the gyromagnetic ratio. The combined effects of dissipation and a spin polarized current are modeled by the term $\frac{\mathbf{m}}{M_s} \times \left(\beta \mathbf{h}^{true} - I\mathbf{h}^J\right)$, where $\beta$ is a parameter governing the dissipation, $\mathbf{h}^J$ is the polarization of the current, $I$ is an empirical factor measuring the strength of the coupling (in units of magnetic field where 1000 Oe corresponds to $10^8$ A/cm$^2$; in the first approximation we

Manuscript received November 2, 2006. The work was sponsored by DOE Grant DE-FG02-06ER46278.

K. Rivkin is with Texas A&M University, Department of Physics, 4242 TAMU, College Station, TX 77843 (phone: (979) 739-0103, e-mail: rivkin@rkmag.com).

J. B. Ketterson, is with Northwestern University, Tech F219, Evanston Campus, Department of Physics and Astronomy, 2145 Sheridan Road Evanston, Illinois 60208 (e-mail: j-ketterson@northwestern.edu).

W. Saslow is with Texas A&M University, Department of Physics, 4242 TAMU, College Station, TX 77843 (e-mail: wsaslow@tamu.edu).

neglect its dependency on the angle between $\mathbf{m}$ and $\mathbf{h}^J$) and $\mathbf{h}^{true}$ is the magnetic field in the sample arising from external sources and demagnetization effects. Our coordinate system is chosen such that the easy axis is parallel to z with the magnetic layers lying in the y-z plane. The current direction is along x and it is polarized along z. In our case a current of 1 mA should correspond to $I \cong 144$ Oe, and the damping coefficient $\beta$ is taken as 0.014[5]. Here we neglect spin pumping, effect which to some extent is accounted for by using an experimentally obtained value of $\beta$.

We are interested in the excitation of a uniform spin wave in the vicinity of the equilibrium magnetization $\mathbf{m}^{(0)}$. Approximating the magnetization of the free layer as a single magnetic dipole, and linearizing Eq. (1) with respect to a small time-dependent departure from the equilibrium direction, $\mathbf{m}^{(1)}(t) = \mathbf{m}^{(1)} e^{-i\omega t}$, and corresponding oscillating magnetic fields, $\mathbf{h}^{(1)}(t)$, one obtains:

$$\frac{d\mathbf{m}^{(1)}}{dt} + \gamma\left[\mathbf{m}^{(1)} \times \mathbf{h}^{(0)} + \mathbf{m}^{(0)} \times \mathbf{h}^{(1)}\right] + \frac{\gamma\beta}{M_s}\mathbf{m}^{(0)} \times \left[\mathbf{m}^{(0)} \times \mathbf{h}^{(1)} + \mathbf{m}^{(1)} \times \mathbf{h}^{(0)}\right] \approx -\gamma \mathbf{g}(t) \quad (3)$$

$$\mathbf{g}(t) = -\frac{I}{M_s}\mathbf{m}^{(0)} \times (\mathbf{m}^{(0)} \times \mathbf{h}^J) - \frac{I}{M_s}\mathbf{m}^{(0)} \times (\mathbf{m}^{(1)} \times \mathbf{h}^J);$$

other terms, like $\frac{I}{M_s}\mathbf{m}^{(1)} \times (\mathbf{m}^{(0)} \times \mathbf{h}^J)$, are significantly smaller ($\mathbf{m}^{(0)} \times \mathbf{h}^J \approx 0$). This equation can easily be solved in two steps: first we find the eigenvalue $\omega$ (i.e. the resonant frequency of small angle oscillations) and eigenvector $\mathbf{V}$ (which describes the orbit of small oscillations) of the homogeneous equation:

$$-i\omega\mathbf{V} + \gamma\left[\mathbf{V} \times \mathbf{h}^{(0)} + \mathbf{m}^{(0)} \times \mathbf{h}^{(1)}(\mathbf{V})\right] + \frac{\gamma\beta}{M_s}\mathbf{m}^{(0)} \times \left[\mathbf{m}^{(0)} \times \mathbf{h}^{(1)}(\mathbf{V}) + \mathbf{m}^{(1)} \times \mathbf{h}^{(0)}\right] = 0$$
(4)

Note that $\mathbf{m}^{(0)} \times (\mathbf{m}^{(0)} \times \mathbf{h}^J)$ is time-independent; it can be shown to be prevalent for very small values of the applied current and corresponds to a small shift in the equilibrium configuration $\mathbf{m}^{(0)}$. We will primarily be interested in the term $\mathbf{m}^{(0)} \times (\mathbf{m}^{(1)} \times \mathbf{h}^J)$; we will analyze the excitation of a uniform mode by expanding the solution of Eq. (3) in terms of the homogeneous solution Eq. (4), which can be done since such solutions form a complete basis set:

$$\mathbf{m}^{(1)} = a(t)\mathbf{V} \cdot \quad (5)$$

Eq.(3) then becomes:

$$\frac{da(t)}{dt} + ia(t)\omega = \gamma\frac{I}{M_s}a(t)\left(\mathbf{m}^{(0)} \times (\mathbf{V} \times \mathbf{h}^J)\right) \cdot \mathbf{V}_L^* \quad (6)$$

where $\mathbf{V}_L^*$ is a left eigenvector of Eq.(4). Since Eq.(4) does not result in a Hermitian matrix, right eigenvectors of Eq.(4) are only orthogonal with respect to $\mathbf{V}_L^*$, left eigenvector of Eq.(4). Eq.(6) has a simple solution, assuming that the polarization is along z direction:

$$a(t) = a(0)e^{-i\omega t}e^{\gamma I \frac{m_z^{(0)}}{M_s} t}. \quad (7)$$

Since $\omega = \omega' - i\omega''$ where $\omega''$ is due to the damping term in the Landau-Lifshitz equation, we can rewrite Eq. 7 as

$$a(t) = a(0)e^{-i\omega' t}e^{\left(\gamma I \frac{m_z^{(0)}}{M_s} - \omega''\right)t}. \quad (8)$$

Note there are two regimes: i) if $\gamma I \frac{m_z^{(0)}}{M_s} < \omega''$, the oscillations die out exponentially in time, and ii) if $\gamma I \frac{m_z^{(0)}}{M_s} > \omega''$, the excitation of the uniform mode increases exponentially. As first predicted by Li and Zhang[10] the initial exponential expansion can result in an oscillatory behavior: for values of $m_z^{(0)}$ close to $M_s$ the following will be satisfied: $\gamma I \frac{m_z^{(0)}}{M_s} - \omega''(m_z^{(0)}) = 0$. To describe the dependence of $\omega''$ on $m_z^{(0)}$ we can use a Kittel formula modified to account for the damping[12]:

$$\omega = \gamma\sqrt{\left(H_0 + m_z^{(0)}B\right)\left(H_0 + m_z^{(0)}A\right)} - i\gamma\frac{\beta}{M_s}m_z^{(0)}\left(\frac{\left(H_0 + m_z^{(0)}B\right) + \left(H_0 + m_z^{(0)}A\right)}{2}\right) \quad (9)$$

$A \equiv (A_{zz} - A_{yy})$
$B \equiv (A_{zz} - A_{xx})$
$C \equiv (A_{yy} - A_{xx})$

where $H_0$ is the external d.c. field applied to the layer and $A_{\alpha\beta}$ is a demagnetization tensor that gives an average demagnetization field inside the layer:

$$H_\alpha = \sum_\beta A_{\alpha\beta}M_\beta. \quad (10)$$

For the above-mentioned ellipse ($130 \times 70$ nm), the numerically-estimated principal elements are $A_{zz} = -0.27$, $A_{yy} = -0.61$, and $A_{xx} = -11.69$. The steady oscillations than occur when:

$$\gamma I \frac{m_z^{(0)}}{M_s} = \gamma\frac{\beta}{M_s}m_z^{(0)}\left(\frac{\left(H_0 + m_z^{(0)}B\right) + \left(H_0 + m_z^{(0)}A\right)}{2}\right). \quad (11)$$

$$\frac{2}{\beta}\frac{I - \beta H_0}{A + B} = m_z^{(0)}$$

Such dynamic equilibrium becomes unstable for sufficiently high current values, resulting in switching of the magnetization.

This analysis tells us that for the currents above the threshold given by Eq.(8), but below the values for which nonlinear effects and the excitations of non-uniform spin waves[13] become important (typically currents above 3 mA[14]), the spin valve radiates a magnetic field having a complicated time-dependence, but which will approximately include the resonant frequency of the uniform mode $\omega'$. Previously we showed that if one applies an r.f. pulse with a time dependent frequency that always matches the resonant frequency, magnetization reversal can occur[3] (the analysis was given for systems with uniaxial anisotropy where resonant frequency depends on the angle between the magnetization and the easy axis). One might then ask whether the magnetic field produced by the excitation of the uniform mode, as described by Eq.(8), can be used to facilitate the magnetization reversal in *another* magnetic nanodot. This would require that both magnetizations rotate almost simultaneously, staying in phase with one another.

**Figure 1 here.**

In order to test this idea we constructed the following numerical experiment. We considered two identical 130 by 70nm ellipses with the material parameters discussed above (Figure 1). The first ellipse is approximated as a uniformly magnetized sample (i.e. the macrospin approximation) that is subjected to the spin transfer torque and average magnetic field arising from the second ellipse. This second ellipse is treated micromagnetically as a collection of approximately 1000 uniformly magnetized prisms; it is subject to the internal dipole-dipole and exchange fields and external magnetic field of the first layer. We then attempted to reverse the magnetization of both ellipses by applying a non-zero current to the first ellipse. Initially the samples are magnetized in the opposite directions along the easy axis (semi-major axis of the ellipse). One can also use other configurations, such as two samples magnetized in the same direction, however such geometries are less effective. A small random perturbation is applied to simulate finite temperature effects. As a result of these simulations we were able to establish: i) magnetization reversal is possible only when these layers are both lying in y-z plane on top of each other, separated by a distance not exceeding 8nm between their surfaces; ii) the precession of magnetization in the second ellipse is uniform, which partially justifies the macrospin approximation chosen by us for the first ellipse; and iii) the magnetization reversal occurs in times similar to those predicted by our theory of resonant switching.

**Figure 2 here.**

In Figure 2 we show the dependence of the magnetization switching time on the values of the applied current; here we defined the switching time as that required for the magnetization to reach the value $m_z^{(0)} = 0$, since after this one can rely on damping alone to complete the reversal (alternatively one can simply leave the current on until the reversal is completed). By itself, the first ellipse switches at the current of 0.76 mA; approximately twice that current is required to switch both ellipses. The switching time depends on the applied current; however a limit is reached at approximately 2.3 mA after which the magnetization of the first ellipses precesses faster than that of the second one. For a sufficiently large current the two ellipses are so far out of phase with respect to one another, i.e. the switching of the first sample due to the applied current occurs much faster than it would be possible to switch the second sample via the generated r.f. field, whose magnitude (and therefore the shortest switching time of the second sample) does not depend on the current, that switching no longer occurs. In our analysis we neglected the non-uniform magnetization of the first ellipse, however the simulation shows that the field inside this ellipse remains essentially uniform, justifying our use of a macrospin approximation.

In order to simulate an actual device that incorporates the phenomena described here one needs to account for the fields due to the fixed layer of the spin valve, which can to some extent be compensated by external d.c. fields. Assuming this and other issuers are addressed, one might indeed use the r.f. radiation produced by a spin valve to facilitate a fast magnetization reversal in a magnetic nanodot, a technology that can potential lead to magnetic storage or magnetic logical elements.

GE-02 4

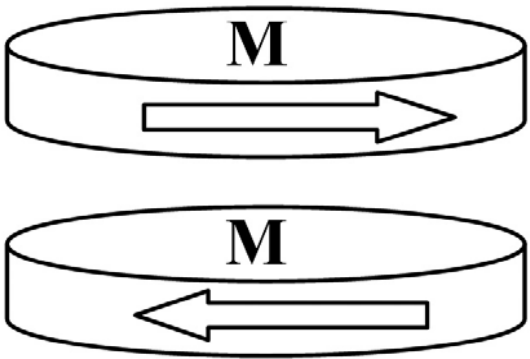

Figure 1. Spin valve's free layer (upper element) and the magnetic nanoparticle (lower element).

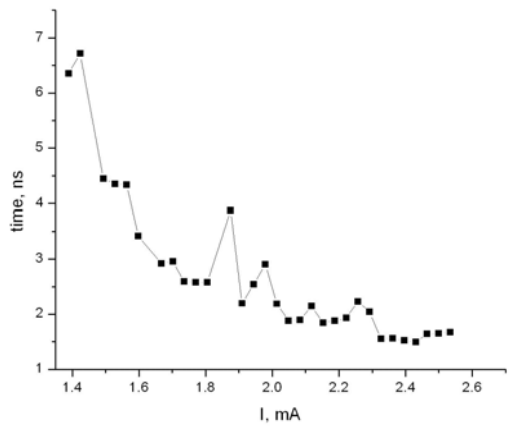

Figure 2. Magnetization switching time as a function of the applied current.